\documentstyle[aaspp4,epsfig,tighten,flushrt]{article}
\def\simlt{\lower.5ex\hbox{$\; \buildrel < \over \sim \;$}}
\def\simgt{\lower.5ex\hbox{$\; \buildrel > \over \sim \;$}}
\def\mag{\mbox{ mag}}
\def\kms{\mbox{ km s$^{-1}$}}

\def\kpc{\mbox{ kpc}}
\def\pc{\mbox{ pc}}

\def\yr{\mbox{ yr}}
\def\msun{\mbox{ M}_\odot}
\newcommand\beq{\begin{equation}}
\newcommand\eeq{\end{equation}}
\newcommand{\etal}{et~al.\ }

\begin{document}

\title{Compound gravitational lensing as a probe of dark matter substructure within galaxy halos}

\author{R. Benton Metcalf}
\affil{\it Institute of Astronomy, University of Cambridge, Cambridge CB3 0HA, UK} 
\and 
\author{Piero Madau}
\affil{\it Department of Astronomy and Astrophysics, University of California,
Santa Cruz, CA 95064}

\begin{abstract}
We show how observations of multiply--imaged quasars at high redshift 
can be used as a probe of dark matter 
clumps (subhalos with masses $\simlt 10^9\msun$) within the virialized extent
of more massive lensing halos. A large abundance of such satellites 
is predicted by numerical simulations of galaxy formation in cold dark 
matter (CDM) cosmogonies. Small--scale structure within galaxy halos affects 
the flux ratios of the images without appreciably changing their positions.
We use numerical simulations to quantify the effect of dark matter 
substructure on the distribution of magnifications, and  
find that the magnification ratio of a typical image pair will deviate 
significantly from the value predicted by a smooth lensing potential if,
near the Einstein radius, only a few percent of the lens surface density is
contained in subhalos.  The angular size of the 
continuum source dictates the range of subclump masses that can have a 
detectable effect: to avoid confusion with gravitational microlensing
caused by stars in the lens galaxy, the background source must be larger
than the optical continuum--emitting region of a QSO.  
We also find that substructure will cause distortions to images on
milli--arcsecond scales and bias the distribution of QSO magnification
ratios -- two other possible methods of detection.

\end{abstract}

\section{Introduction}

The popular model of hierarchical structure formation in a universe 
dominated by cold dark matter (CDM), while quite successful in matching the
observed large--scale density distribution, is currently facing a 
`small--scale crisis' (e.g., Moore 2001). First and foremost, CDM simulations 
appear to produce halos that are too centrally concentrated
compared to the mass distribution inferred from the rotation curves of 
(dark matter--dominated) dwarf galaxies \markcite{1999MNRAS.310.1147M}
({Moore} {et~al.} 1999b; Klypin \etal 2001; Navarro, Frenk, \& White 1997). 
In addition to the cuspy density profiles, the predicted abundance of 
satellites of mass greater than $10^8\,\msun$ within the virialized 
extent of a Milky Way's halo is more than an order of magnitude larger than 
the number of dwarf galaxies with comparable mass observed within the Local 
Group \markcite{1999ApJ...524L..19M,
1999ApJ...  522...82K}({Moore} {et~al.} 1999a; {Klypin} {et~al.} 1999b; Mateo
1998). A number of related but more complex problems is also being 
discussed in the literature (see Sellwood \& Kosowsky 2001 for a recent
review), together with the significance of some of the observations 
(van den Bosch \& Swaters 2001). 
It is unclear at this stage whether the CDM paradigm is actually 
incorrect and needs to be modified to include, e.g., self--interacting
(Spergel \& Steinhardt 2000) or warm dark matter (Bode,
Ostriker, \& Turok 2000; Colin, Avila--Reese, \& Valenzuela 2000), or whether 
some of these discrepancies may have a more `astrophysical' 
origin. Feedback processes such as photoionization, for example, may 
prevent gas cooling and inhibit star formation in the majority of the 
dwarf subhalos (e.g. Bullock, Kravtsov, \& Weinberg 2001).
In this case the dark matter satellites of the Milky Way would still be
present, but only a small fraction of them would actually be visible,
having formed stars prior to the reheating of the intergalactic medium.

Gravitational lensing is a powerful tool for studying the structure
of dark matter halos. The statistics of wide--separation lenses
can be used to constrain the amount of mass in the cores of dark matter
halos on group and cluster mass scales (Keeton \& Madau 2001; Li \& Ostriker
2001; Flores \& Primack 1996). Similarly, substructure in galaxy halos can 
be probed by gravitational lensing even if the satellites contain too few
baryons to be observable. A lump approximating a singular isothermal
sphere (SIS) of one--dimensional velocity dispersion $\sigma=10\,\kms$ 
(corresponding to a mass of order $10^8\msun$) produces a deflection angle of 
about 3 milli--arcsecond if it acts as a lone lens.  If the lump resides in a
larger halo this deflection will effectively be magnified.  This can
cause distortions of images on ten milli--arcsecond scales and
appreciable changes to the total flux.

It has been pointed out by Mao \& Schneider (1998) that substructure 
in the lens galaxy can explain the discrepancy between the observed and
(`simple') model--predicted flux ratios in the quadruply--imaged QSO 
B1422$+$231. These authors considered two illustrative pictures of
substructure, point--like ``globular clusters'' with masses of $10^6\,\msun$ and 
a plane--wave smooth fluctuation. In this paper we study how to use 
gravitationally lensed QSOs as a probe of the existence of dark matter 
subclumps within the lensing potential, the small--scale structure that is 
indeed predicted by numerical CDM simulations. The format 
is as follows.  In \S~\ref{sec:Surv-Substr-Galaxy} we discuss 
the survival of subclumps in dark matter halos.  A model for halo
substructure is presented in \S~\ref{sec:Subcl-Distr} where it is argued
that many subclumps will survive to within a few kpc of a galactic
halo's center.  The formalism for describing lensings by subclumps is
introduced in \S~\ref{sec:Lensing-substructure}, and the numerical
simulations used to study the magnification and image distribution are
described in \S~\ref{sec:Simulations}.  Specific examples of lensing
systems and subclump distributions are investigated in
\S~\ref{sec:Examples}.  Finally, we discuss the implications of this
work in \S~\ref{sec:Discussion}.
\section{Survival of substructure in galaxy halos}
\label{sec:Surv-Substr-Galaxy}

In hierarchical cosmogonies -- of which CDM is the best studied example --
all the mass of the universe at early times is contained in small--mass halos 
that later merge into larger and larger systems.
The problem of how ``halos orbiting within halos'' evolve with time is 
rather complex \markcite{1995ApJ...451..598J,1996ApJ...457..455M,1999ApJ...516..530K}({Johnston}, {Spergel}, \&  {Hernquist} 1995; {Moore}, {Katz}, 
\&  {Lake} 1996; {Klypin} {et~al.} 1999a). Survival of satellites in 
galaxy clusters has been investigated by a
number of authors \markcite{1998MNRAS.299..728T,1998MNRAS.300..146G}(see {Tormen}, {Diaferio}, \&  {Syer} 1998; {Ghigna} {et~al.} 1998)
where it has been found that somewhere between $10\%$ and $20\%$ of the
cluster mass remains in bound subclumps. Simulating substructure in
galaxy halos is much more computationally demanding.
Very high resolution cosmological N--body simulations
\markcite{1999ApJ...522...82K,1999ApJ...524L..19M}({Klypin} {et~al.} 1999b; {Moore} {et~al.} 1999a) and semi--analytic
modeling \markcite{2000ApJ...539..517B}({Bullock}, {Kravtsov}, \&  
{Weinberg} 2000) have shown that a large number of subclumps do survive in 
CDM theories, and that galaxy halos resemble scaled down versions of galaxy 
clusters. According to these simulations,
the mass function of subclumps is approximately $dn/dm\propto m^{-2}$ down to
a typical resolved mass of $\simgt 10^7\msun$.

\begin{figure}[t]
\centering\epsfig{figure=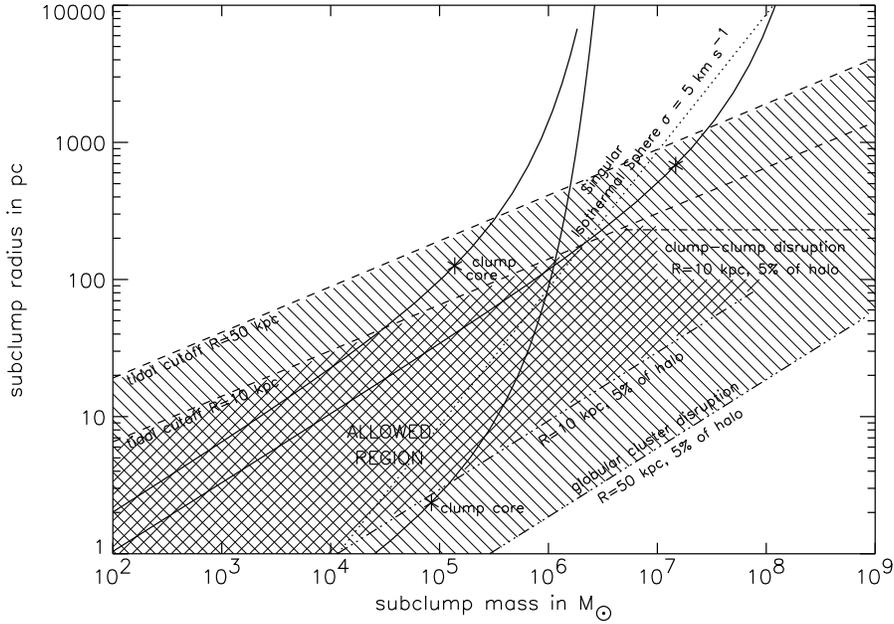,height=3.5in}
\caption[qiw]{\footnotesize Survival of dark matter satellites 
in a host halo modeled as a singular isothermal sphere of circular 
velocity $V_c=226\kms$. {\it Dashed lines:} tidal stripping radius as a 
function of subclump mass at two different distances from the host halo 
center, $R=10$ and $50\kpc$. {\it Dot--dot--dashed lines:} constraints 
imposed by the survival of globular clusters of mass
$m_g=5\times10^{4}\msun$ and radius $r_g=10\pc$ for 10~Gyr.
{\it Dot--dashed line}: collisional timescale between
subclumps, assuming $t_{\rm ss}=10$~Gyr (see text for details).
{\it Dotted curve:} $m(r)$ for a SIS with $\sigma=5\kms$.  
{\it Solid curves:} three examples of $m(r)$ for NFW halos which differ in the
choice of $r_s$ and $\delta_c$.  The asterisks mark $r_s$.  The two
curves with larger $r_s$  follow the $r_s$--$\delta_c$
relation given in \markcite{1997ApJ...490..493N}{Navarro} {et~al.}
(1997) for the $\Omega_M=1$ SCDM model with $h=0.65$ extrapolated to these
masses.  In a $\Lambda$CDM model with $\Omega_\Lambda=0.75$, $r_s$ is
about $0.6$ times smaller for the same clump mass, so satellites will penetrate
further into the halo. 
}
\label{fig:tidal_stripping}
\end{figure}

To understand the destruction process of dark matter clumps and estimate the 
range of substructure masses and radii that are to be expected, we can use 
some simple 
analytic arguments.  Dynamical friction will cause only the most massive 
satellites, $m\simgt 10^{10}\msun$, to sink into the center of galaxy size 
halos.
For the range of (smaller) masses we are interested in, the most important 
factor in the evolution of substructure is tidal stripping within the 
parent halo that they inhabit. The tidal radius $r_t$ of a clump of bound
mass $m$ can be estimated as 
\begin{eqnarray}\label{tidal_radius}
r_t\simeq R \left[{m\over 3 M(R)}\right]^{1/3}
\end{eqnarray}
where $R$ is measured
from the center of the (host) halo, and $M(R)$ is the halo mass within that
radius \markcite{BinneyAndTremaine}(Binney \& Tremaine 1987). This 
constraint on the subclump size is plotted in 
Figure~\ref{fig:tidal_stripping} for a Milky Way--like galaxy halo
approximating a SIS mass density profile,
\beq
\rho(r)={\sigma^2_{\rm halo}\over 2\pi G r^2},
\eeq
with circular velocity $V_c=\sqrt{2}\sigma_{\rm halo}=226\,\kms$, and for 
two different radii $R$. Any clump of radius $r>r_t$ will be 
stripped of mass as it falls towards the center. By losing mass and shrinking
in size the clump may eventually become quasistable when it reaches
approximately the tidal stripping line evaluated at the pericenter of
its orbit (dynamical friction will eventually drag them in on a long
timescale and gravitational heating caused by the time varying tidal
force will slowly pump energy into them).  Four such tracks are shown in
Figure~\ref{fig:tidal_stripping} under the assumption that the clump
density profile remains unaffected by mass stripping.

SIS subclumps will lose a large fraction of their mass before they can
reach the solar radius, $R\simeq 10\kpc$.  A clump with a more realistic
NFW profile \markcite{1997ApJ...490..493N}({Navarro},  
{Frenk}, \&  {White} 1997),
\beq
\rho(r)={\rho_c\delta_c\over (r/r_s)(1+r/r_s)^2}, \label{nfw_profile}
\eeq
is more centrally concentrated and can reduce its size without loosing 
much of its mass.  This is only true until $r_t(R)\simlt r_s$, at which
point the satellite looses its mass more rapidly.  Here $r_s$ is the
subclump scale size, i.e. the radius where the logarithmic slope of the
profile is $-2$, $\rho_c\equiv 3H^2/8\pi G$ is the critical density, and
$\delta_c$ indicates the characteristic density contrast of the clump. In Figure~\ref{fig:tidal_stripping} 
the radii $r_s$ for three NFW tracks are
marked with asterisks. If the core of the NFW profile is well below the
tidal stripping boundary the subclump can survive in the inner halo
without losing much mass.  According to the $r_s$--$\delta_c$ relation
given in \markcite{1997ApJ...490..493N}{Navarro} {et~al.} (1997), 
extrapolated to the mass range relevant for this paper (two examples are
shown in the figure), SCDM halos are right near this boundary of
destruction and survival at $R\sim 30\kpc$.
High--resolution N--body simulations show, however, a large scatter about 
this relation \markcite{astro-ph/9908159}(Bullock {et~al.} 1999) and in 
low density models such as currently popular $\Lambda$CDM the halos are
more concentrated and thus more able to survive -- $r_s$ is reduced by a
factor of $\sim 0.6$.
Since it takes several orbits for a subclump to be completely
destroyed even if unstable to tidal disruption, and satellites are being
continuously accreted by more massive galaxies, it is expected that a
significant population of subclumps will exist within the Einstein
radius of the host halo,
\beq
\theta_E D_l=4\pi\left({\sigma_{\rm halo}\over c}\right)^2\,{D_{\rm 
ls}D_l\over D_s}
\approx 10\,h^{-1}\,{\rm kpc} {H_0 D_{\rm ls}D_l\over c D_s} 
\left({\sigma_{\rm halo}\over 156\,\kms}\right)^2.
\eeq
Here $D_l$, $D_s$, and $D_{\rm ls}$ are the angular size distances to the 
lens, source, and from the lens to the source, respectively, and
$H_0=100\,h\,\kms$ Mpc$^{-1}$ is the Hubble constant today.  The
fraction of mass that remains in subclumps is investigated more fully in
section~\ref{sec:Subcl-Distr} where a specific substructure model is adopted.

Another constraint on subclumps comes from the requirement that they not 
disrupt the observed population of globular clusters.  This is actually more 
stringent than the constraint imposed by over heating of the Galactic 
disk, which requires a large density of $m > 10^6\msun$ satellites
\markcite{1987ApJ...316...23C,1995ApJ...442L...5M}(Carr \& Lacey 1987; Moore 
\& Silk 1995).  If the subclumps are larger than globular clusters, 
the disruption timescale $t_{\rm sg}$ can be estimated in the same way as
\markcite{1958ApJ...127...17S}{Spitzer} (1958) estimated the timescale for 
disrupting open star clusters by molecular clouds 
\markcite{BinneyAndTremaine}(Binney \& Tremaine 1987), i.e.
\begin{eqnarray}\label{eq:globular}
t_{\rm sg}=\frac{E}{\dot{E}} \sim \frac{0.03 \sigma_{\rm halo} m_g 
r^2}{G r_g^3 m^2 \eta},
\end{eqnarray}
where $\eta$ is the number density of satellites, $\sigma_{\rm halo}$ is the 
velocity dispersion in the parent halo, and $m_g$ and $r_g$ are the mass and
radius of the clusters.  This constraint is shown in 
Figure~\ref{fig:tidal_stripping} assuming that the fraction of the 
halo mass in satellites is $5\%$, $t_{\rm sg}=10$~Gyr, $m_g=5\times10^4\,
\msun$, and $r_g=10\,$pc. It sets a limit on the number of dense clumps,
not on their size or mass.

Similarly, the subclump--subclump collisional disruption timescale can be 
estimated by replacing in equation~(\ref{eq:globular}) the mass and radius 
of a typical globular cluster with those of dark matter substructure, 
\begin{eqnarray}
t_{\rm ss} \sim 0.03\sigma_{\rm halo} \left( Gmr\eta\right)^{-1}.
\end{eqnarray}
This is also plotted in Figure~\ref{fig:tidal_stripping} for
$t_{\rm ss}=10$~Gyr. Again, the subclumps do not really need to survive
this long. Since cosmological simulations show a continuous replenishment of
satellites from infall, their number could remain high even when
collisional disruption is actually taking place.
The survival of the baryonic content of infalling clumps is further
complicated by shock stripping. Most of the gas may be removed in this way,
making it difficult to interpret any observations of visible matter in 
terms of dark satellites, for example in the case of the high velocity halo
clouds detected in neutral hydrogen \markcite{1999ApJ...514..818B}({Blitz} 
{et~al.} 1999).

\section{Distribution of satellites}
\label{sec:Subcl-Distr}

We seek to develop a simple model for satellites within host halos 
that incorporates all the aspects of real substructure that are
important to lensing.  We will use a simple power-law model for the
subclumps' internal structure
\begin{eqnarray}
m(m_a)=m_a (r/r_a)^\beta.
\end{eqnarray}
The parameter $r_a$ is kept fixed while the mass within this distance,
$m_a$, changes for different clumps.  Different stages in the stripping
of an NFW halo can then be represented by different $\beta$'s -- $\beta=2$
for $r_t(R) \simlt r$ and $\beta\simeq0.01$ for $r_t(R) \simgt r$.  For
the special case of a SIS subclump $\beta=1$ and $m_a=2\sigma^2 r_a/G$.
Combining this with the tidal radius~(\ref{tidal_radius}) gives the mass
as a function of galactic radius 
\begin{eqnarray}
m(m_a,R)=\left[ \frac{m_a^{3/\beta}}{3 M(R)} \left( \frac{R}{r_a} \right)^3
\right]^{\frac{\beta}{3-\beta}}. \label{eq:mass_general}
\end{eqnarray}

In our model the number density of subclumps, $\eta$, is assumed to be
proportional to the average or total mass density -- clumped plus smooth
components.  In this way the clumps flow with the mass.  Dynamical
friction is taken to be unimportant.  The distribution of satellites is
modeled as a power-law in $m_a$,
\begin{eqnarray} \label{eq:mass_func}
\frac{1}{\overline{\rho}}\frac{d\eta}{dm_a}(m_a)=\frac{G}{4m_0
r_a c^2}
\left( \frac{m_a}{m_a^{\rm max}} \right)^{\frac{\beta-3n}{3-\beta}},
\end{eqnarray}
where $\overline{\rho}$ is the total local density and $m_a^{\rm max}$
is the maximum normalized mass.  The above is an $R$-independent quantity.  The exponent is chosen so that the true
mass function is $d\eta/dm \propto m^{-n}$ where the normalization will
depend on $R$ and the structure of the host halo.  CDM simulations favor
$n \simeq 2$.

The minimum normalized mass will be set by the parameter
$f_m\equiv m_a^{\rm min}/m_a^{\rm max}$: the requirement that the
mass in clumps cannot exceed the total halo mass sets a lower bound on
$f_m$ if the mass function is steep enough.  There is a mass below
which satellites do not contribute significantly to the lensing and this
cutoff is in practice more restrictive than the former bound.  For this
reason the lensing properties become insensitive to $f_m$ when it
is below some level.  

\begin{figure}[t]
\centering\epsfig{figure=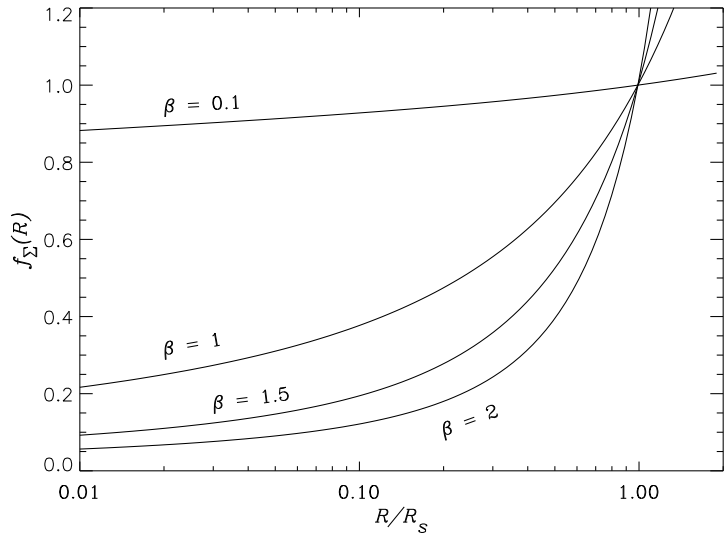,height=3.0in}
\caption[qiw]{\footnotesize The fraction of surface density in
substructure as a function of radius from the center of a NFW halo.  The
scale size of the host halo is $r_s=R_S$ (see
equation~\ref{nfw_profile}).  This is independent of $\delta_c$.  In all
cases the normalization is chosen so $f_\Sigma(R_s)=1$.  SIS subclumps
are labeled $\beta=1$.  The core of NFW profiles is
$\beta=2$ and the core of a Moore profile (Moore \etal 1999b) is 
$\beta=1.5$.  For comparison $R_s$ is expected to be $\sim 20\kpc$ for
the Milky Way halo.}
\label{fig:f_Sigma}
\end{figure}

With equations~(\ref{eq:mass_general}) and~(\ref{eq:mass_func}) we can
easily calculate the mass fraction in subclumps as a function of
galactic radius for fixed $\beta$.  Integrating this along the line of
sight gives the fraction of surface density in substructure,
$f_\Sigma(R)$.  This has been done in figure~\ref{fig:f_Sigma} for a NFW
host halo.  It can be seen here that for small $\beta$, which
approximates the case of NFW subclumps with $r_s\simlt r_t(R)$,
$f_\Sigma(R)$ remains relatively independent of $R$.  By the definition
of a subclump 
$r_s < R_s$ so we expect that substructure will survive relatively intact
in the region $R \simgt R_s$.  As $R$ further decreases $f_\Sigma(R)$
will slowly decrees until $r_t \simeq r_s$ at which point it will drop
more steeply.  It seems realistic that at $R\sim 1 \kpc$, the
scale on which multiple QSO images are formed, $f_\Sigma(R)$ does not drop
below $20\%$ of the what it is at $R=R_s$ ($\sim 20\kpc$ for the Milky
Way).  Numerical simulations find that $10 - 20\%$ of the mass within the virial radius
of a $10^{12}\msun$ halo is contained in substructures of 
mass $> 10^7\msun$ (Moore \etal 1999a).  The lensing properties will be
sensitive to structures that are several orders of magnitude smaller --
below the resolution of any existing simulation.  This simple model
strongly supports our conjecture that in the CDM model $f_\Sigma(R)$ is
greater than a few percent at the point where multiple images are formed and
could potentially be much larger.  An accurate determination of
$f_\Sigma(R)$ for the relevant mass range will require CDM simulations
with mass resolution $4-5$ orders of magnitude better then those
available today.

For the lensing simulations discussed later the subclumps will be modeled
as SISs and the distribution will be renormalized to the desired
$f_\Sigma$.  The distributions of magnifications and image distortions
will be only weakly dependent on the SIS assumption because the
angular sizes of the sources we consider are not small compared to the
sizes of the subclumps, as will be seen in
section~\ref{sec:Image-distortions}.  In this case the magnification
averaged over the image will not depend on the details of the
subclumps' internal structure.  The lensing is more dependent on
$f_\Sigma$ which makes it a more useful free parameter.
For SIS subclumps $m_a$ can be replaced by $\sigma$ in the above giving
\begin{eqnarray}\label{r_m_general}
r_t(\sigma,R)= \sigma \sqrt{ \frac{2R^3}{3GM(R)} } ~~~~~
m(\sigma,R)=\sigma^3 \sqrt{ \frac{8R^3}{3G^3M(R)} }
\end{eqnarray}
\begin{eqnarray}
\frac{1}{\overline{\rho}}\frac{d\eta}{d\sigma}(\sigma)=\frac{1}{m_0
\sigma_{\rm max}}
\left( \frac{\sigma}{\sigma_{\rm max}} \right)^{2-3n}.
\end{eqnarray}

\section{Lensing by substructure}
\label{sec:Lensing-substructure}

A probe of purely gravitational mass is needed to confirm 
or disprove the existence of substructure in galaxy halos.
We are interested in the situation where there is one galaxy size
gravitational lens creating multiple images of a background source.
Within this lens there is a large number of randomly positioned
subclumps that affect each of the images.  We will ignore the effects of
any additional clumps outside of the main lens.  The large number
of subclumps and the small, but not vanishingly small, size of the source
will require simulations of a small region of the lens plane rather than
the entire lens at once.  The contribution from the parts of the lens outside 
the simulation region must therefore be represented in some way.
The theoretical background to our method is presented in this section.

If no substructure existed in the lens it would have the surface density
$\kappa_{\rm smooth}(\vec{x}')$ expressed in units of the critical
density, $\Sigma_c\equiv c^2 D_s(4\pi G D_lD_{\rm ls})^{-1}$.  We will
subdivide this density into two components within the region of the lens
plane being simulated; the substructure component, $\kappa_{\rm
sub}(\vec{x}')$, and the halo component,
\begin{eqnarray}
\kappa_{\rm halo}(\vec{x}') =
\left\{ \begin{array}{lcc}
\kappa_{\rm smooth}(\vec{x}') &,& {\rm outside~region} \\
\left[1- f_\Sigma(\vec{x}_o)\right] \kappa_{\rm
smooth}(\vec{x}') &,& {\rm inside~region}
\end{array} \right.
\end{eqnarray}
where $\vec{x}_o$ is the center of the region.  The simulation region is
made large enough that, for the purposes of lensing, the mass outside the
region can be considered smoothly distributed.  It is assumed that
$f_\Sigma(\vec{x}')$ does not change over the simulation region.

In the thin lens approximation, the light deflection angle caused by the
two surface density components will add.  This follows from
linear superposition of the surface potentials.  The lensing equation
can thus be written
\begin{eqnarray}\label{orig_lens_eq}
\vec{y}'=\vec{x}'-\vec{\alpha}_{\rm halo}(\vec{x}')-\vec{\alpha}_{\rm sub}(\vec{x}').
\end{eqnarray}
Here the unlensed angular position of the source,
$\vec{\theta_s}$, the position of the image, $\vec{\theta}_i$,
and the deflection angle, $\vec{\alpha}'(\vec{x})$, have been rescaled
\begin{eqnarray}\label{scalings}
\vec{x}'=\frac{\vec{\theta_i}D_l}{\lambda_o}
~~~~~~\vec{y}'=\frac{\vec{\theta_s}D_l}{\lambda_o} 
~~~~~~\vec{\alpha}(\vec{x}')=\frac{D_l D_{ls}}{\lambda_o D_s} \vec{\alpha}'(\lambda_o\vec{x})
\end{eqnarray}
where $\lambda_o$ is any arbitrary scale that can be set to a
convenient value (usual making $|\alpha(x)| \sim 1$).
In the absence of substructure the image in question will appear at
$\vec{x}_0$:
\begin{eqnarray}\label{0th_order_lq}
\vec{y}_0=\vec{x}_0-\vec{\alpha}_{\rm smooth}(\vec{x}_0).
\end{eqnarray}
Subtracting equation~(\ref{0th_order_lq}) from~(\ref{orig_lens_eq}), and 
shifting to the coordinates $\vec{x}\equiv \vec{x}'-\vec{x}_0$ and 
$\vec{y}\equiv \vec{y}'-\vec{y}_0$, yields
\begin{eqnarray}
\vec{y}=\vec{x}-\vec{\alpha}_{\rm halo}(\vec{x}+\vec{x}_0)
+\vec{\alpha}_{\rm smooth}(\vec{x}_0)-\vec{\alpha}_{\rm sub}(\vec{x}+\vec{x}_0).
\end{eqnarray}
We need to express $\vec{\alpha}_{\rm halo}(\vec{x})$ in terms
of $\vec{\alpha}_{\rm smooth}(\vec{x})$.  We make the approximation
$\kappa_{\rm smooth}(\vec{x}')-\kappa_{\rm halo}(\vec{x}') = f_\Sigma
\kappa_{\rm smooth}(\vec{x}_o)$ inside the simulation region.  The
constant surface density results in a linear, isotropic deflection term $f_\Sigma
\kappa_{\rm smooth}(\vec{x}_o)\,\vec{x}$ so that the final lensing equation is
\begin{eqnarray}\label{lens:eq}
\vec{y}=\vec{x}-\vec{\alpha}_{\rm smooth}(\vec{x}+\vec{x}_0)
+\vec{\alpha}_{\rm smooth}(\vec{x}_0)+f_\Sigma(\vec{x}_0)\kappa(\vec{x}_0)\,\vec{x}-\vec{\alpha}_{\rm sub}(\vec{x}+\vec{x}_0).
\end{eqnarray}
The magnification matrix, $A_{ij} \equiv \partial
{y^i}/\partial {x^j}$, of equation~(\ref{lens:eq}) is the true observed
magnification matrix.  This is the lensing equation used throughout the
rest of this paper.

One might be concerned that besides the isotropic linear term,
$f_\Sigma(\vec{x}_0)\kappa(\vec{x}_0)\,\vec{x}$, in
equation~(\ref{lens:eq}) there might also be
anisotropic linear, or shear, term of the same order.  This is the
shear caused by the surface density $f_\Sigma \kappa_{\rm
smooth}(\vec{x}+\vec{x}_0)$ inside the simulation region.  If the region
is small compared to the size of the whole halo this shear
turns out to be very small.  To see this consider a spherically symmetric mass
distribution.  The contribution to the shear at point $R$ from the mass
within the shell $R>r>R-\delta r$ is the total shear, $\gamma(r)$, minus 
what the shear would be were the shell empty.  Since $\gamma(r)\propto
r^{-2}$ outside any spherically symmetric mass, and the mass exterior to
$R$ does not contribute to $\gamma(R)$, the contribution of this shell is
\begin{eqnarray}
\gamma_{\rm shell} = \gamma(R) \left[ 1-\frac{(R-\delta r)^2\gamma(R-\delta
r)}{R^2 \gamma(R)} \right].
\end{eqnarray}
For a singular isothermal sphere
$\gamma(r)\propto r^{-1}$, so $\gamma_{\rm shell}/\gamma(R) = \delta r/R$. 
As long as the simulation region is small compared to the distance to the
center of the host halo the mass exterior to this region will dominate
the shear in the smooth halo model.  If the halo is asymmetric it is only
asymmetries on a scale smaller than the simulation region that could change this
conclusion.  In all the cases we will consider, the simulation region
will be a factor of 100 or more smaller than the distance between the
image and the center of the lens, $R$.  We can then safely ignore this
shear term as is done in equation~(\ref{lens:eq}).  Another related
consideration is that, in the limit where the number of subclumps
becomes very large at constant surface density, the distortion of the
image should converge to the smooth lens solution.  In our simulations
the satellites will be distributed uniformly over the region considered
so it is incapable of producing a shear in this limit.  As a result the
extra shear must be zero.

It is instructive at this point to estimate how large of an effect
substructure is expected to have.  Expanding equation~(\ref{lens:eq})
around $\vec{x}_0$ we can estimate the distortion of an image along the two
eigenvectors of the smooth model's magnification matrix,
\begin{eqnarray}\label{eq:lens_estimate}
\theta_I \sim \frac{\theta_s +  \langle \alpha_{\rm sub}\rangle \lambda_o/D_l}{
1-(1-f_\Sigma)\kappa \pm \gamma}.
\end{eqnarray}
It is apparent that the effects of substructure will be magnified or
demagnified by the larger scale lens as represented by the denominator
in equation~(\ref{eq:lens_estimate}). Dark matter subclumps will have 
a significant
effect on the image only when the deflection is a significant fraction
of the angular size of the source $\theta_s \simlt  \langle
\alpha_{\rm sub}\rangle \lambda_o/D_l$.  The question of whether the lensing
effect of satellites is detectable reduces then to the
question of whether deflections of this order are common.  Note that the
probability of a subclump contributing to the lensing is boosted by a
factor equal to the background magnification $\mu^b =
\left([1-(1-f_\Sigma)\kappa]^2 - \gamma^2\right)^{-1}$, which can be
very large near a critical curve where the magnification formally diverges.  
This, combined with the fact that the line of sight already passes
through a high density region, result in substructure having a
larger effect than might otherwise be expected.

\section{Lensing simulations}
\label{sec:Simulations}

To calculate the distribution of magnifications for a large number of random 
configurations of satellites an efficient method of simulating the lensing
effect is required. The basic approach we use is to calculate the
deflection angle, $\vec{\alpha}_{\rm sub}(\vec{x})$, at every point on a
dense grid, and use equation~(\ref{lens:eq}) to determine if $\vec{y}$ lies
within a specified source.  Since surface brightness is conserved the
magnification can be calculated by multiplying the area of a grid cell
by the number of grid points found to be in images and dividing by the
original size of the source.
We are considering images that are 
smaller than the angular size of the host halo, so the grid will cover a small
region over which the average density is approximated as constant.
Multiple sources can be considered simultaneously as long as it is
certain that none of their images are outside of the gridded region.

Direct summation of the contributions to $\vec{\alpha}_{\rm sub}(\vec{x})$ from 
each subclump is ponderously slow (of order $N_{\rm clump}N_{\rm grid}$).  To
speed the calculation up the following approach is used.  We scale all
lengths to a convenient and fixed $\lambda_o$.  After a random
realization of subclump masses and positions is created, 
$\vec{\alpha}_{\rm sub}(\vec{x})$ is calculated in two steps.  
The mass distribution in clumps can be related to the deflection angle
through the lensing potential, which can be broken into two parts
$\psi(x)=\delta\psi(x)+\psi^o(x)$.  The background potential,
$\psi^o(x)$, is the solution to
$\nabla^2\psi^o(\vec{x})=2\,\overline{\kappa}$ where
$\overline{\kappa}$ is the average surface density of substructure
within the simulation region.  This is easily solved.  The perturbed
potential and the deflection angle are then given by
\begin{eqnarray}\label{eq:lens_potential}
\nabla^2\delta\psi(\vec{x})=2\left[\kappa(\vec{x})-\overline{\kappa}\right],~~~~~~~~~~~
\vec{\alpha}_{\rm sub}(\vec{x})=\vec{\nabla}\delta\psi(\vec{x})+\overline{\kappa}~\vec{x}.
\end{eqnarray}
An estimate of $\vec{\alpha}_{\rm sub}(\vec{x})$ can be quickly
found by moving the mass of each clump to its closest grid point and
solving (\ref{eq:lens_potential}) by fast Fourier transform (FFT).

The contribution to
$\vec{\alpha}_{\rm sub}(\vec{x})$ from nearby subclumps is not well represented
in this FFT solution because of the interpolation to a grid and the
internal structure of a subclump.  To refine the calculation a square patch
around each clump is considered.  The contribution of this satellite
to the FFT estimate is subtracted in this patch, and then a more accurate
estimate is recalculated.  Here we use the fact that
the deflection angle for any spherically symmetric lens is particularly
simple \markcite{SEF92}({Schneider}, {Ehlers}, \& {Falco} 1992)
\begin{eqnarray}\label{alpha_of_m}
\vec{\alpha}(x)=\frac{\overline{m}(x)}{x}~\hat{x} ~~~~~~~~~ \overline{m}(x)\equiv 2\int_0^x dx' x'\kappa(x').
\end{eqnarray}
The size of the patch around each satellite is adaptively scaled to be
proportional to the size of the clump.  This approach results in a 
$\sim N_{\rm grid}\log N_{\rm grid} + N_{\rm clump} \overline{N}_{\rm
patch}$ scaling where $\overline{N}_{\rm patch}$ is the average number
of grid points in the patches.  Generally there are more small subclumps
than large ones, and this technique is much faster than the direct sum
over satellites.

In this way the FFT is used to calculate the long range ``forces'', and 
direct summation is used for short range as in a P3M N--body code.  The FFT
results in the effective mass distribution being periodic on the scale
of the box; in practice, however, the mass distribution is always smooth
enough on these scales and the artificial periodicity does not affect
the lensing.

Note that if $\overline{\kappa}=f_\Sigma\kappa_{\rm smooth}(\vec{x}_0)$,
as it is on average, 
the second term in the expression for $\vec{\alpha}_{\rm sub}(\vec{x})$
in~(\ref{eq:lens_potential}) will cancel the fourth term in
equation~(\ref{lens:eq}).  This will generally be the case when there
are large number of subclumps within the simulation region making 
statistical fluctuations small.  This ensures that the results will
converge to the smooth lens case when the subclumps get very dense.  In
the opposite limit where there are no subclumps in the simulation region
$\vec{\alpha}_{\rm sub}(\vec{x})=0$ and equation (\ref{lens:eq}) will give the
correct background magnification.

\section{Examples}
\label{sec:Examples}

A comparison of our simulations with the data requires a specific 
lensing system to be modeled in detail.  An overall smooth lensing
potential needs to be constructed before the effects of substructure on
each image can be investigated.  We will leave a detailed analysis of
specific lensing systems for a later paper, and concern ourselves
here with representative examples that will illustrate what can be
expected.  In section~\ref{sec:SIS-model} some general expectations for
the SIS model that is used in the simulations will be discussed.  Then
in section~\ref{sec:Numerical-results} the results of numerical
simulations are presented and interpreted.

\subsection{SIS model}
\label{sec:SIS-model}

For simplicity, the host halo will be modeled as a SIS.
While to model most observed lens systems some ellipticity or external
shear must be added to the SIS model, the effects of substructure will
not qualitatively change in an asymmetric model.  

Two images will then appear at $x=y\pm 1$ ($y<1$), where the
scale used is
$\lambda_o(\sigma_{\rm halo})=4\pi\sigma_{\rm halo}^2c^{-2} D_lD_{\rm 
ls}/D_s$.  The
unperturbed shear and convergence are 
\begin{eqnarray}
\gamma=\kappa=\frac{\lambda_o}{2 r}=\frac{1}{2 |x|}.
\end{eqnarray}
From equation~(\ref{r_m_general}), the subclump -- also SISs -- have sizes and masses
\begin{eqnarray}\label{sis_m_r}
r(\sigma,R)=\frac{R\sigma}{\sqrt{3}\,\sigma_{\rm halo}} ~~~~~~~ 
 m(\sigma,R)=\frac{2 R\sigma^3}{\sqrt{3}G\sigma_{\rm halo}}.
\end{eqnarray}
The scaled deflection angle caused by a truncated SIS subclump can be
calculated using equation~(\ref{alpha_of_m}),
\begin{eqnarray}
\vec{\alpha}(x)= \frac{2}{\pi}
\left(\frac{\sigma}{\sigma_{\rm max}}\right)^2 \left\{
\begin{array}{cl}
a - \sqrt{ a^2-1 } + \tan^{-1}\sqrt{ a^2-1} &,~~~~ a>1 \\
a &,~~~~ a<1,
\end{array}\right.
\end{eqnarray}
where $a\equiv r(\sigma,R)/x\lambda_o(\sigma_{\rm max})$, and $x$ is the 
distance defined in equation~(\ref{scalings}) using
$\lambda(\sigma_{\rm max})$.

When the impact parameter $y$ is small the images will form at close to
the same $R$ on opposite sides of the lens.  In
section~\ref{sec:Subcl-Distr} it was shown that if the subclumps are
compact enough compared to the host halo their mass distribution will
not be a strong function of $R$.  Following this argument we will take
$f_\Sigma$ to be the same for both images in the simulations.  We will
also replace the mass and radius distributions at a projected distance
$x$ with their values at the radius $R=x\lambda_o(\sigma_{\rm halo})$.
Most of the mass to located near this radius and a more detailed model
would not be warranted at this point.

Let us consider what level of substructure will be important for
lensing in this model.  From \S~\ref{sec:Lensing-substructure}
we know subclumps will have a significant effect on the image when the
deflection caused by an individual satellite is a significant fraction
of the angular size of the source. 
We will use the requirement $\theta_s \simlt 10 \langle 
\alpha_{\rm sub}\rangle \lambda_o/D_l$.  A characteristic deflection angle
is $\langle\alpha_{\rm sub}\rangle \lambda_o \simeq \lambda_o(\sigma)$.  We
can then put a loose lower limit on the mass of subclumps that can have
a significant effect on the lensed image of a source of physical size $l_s$,
\beq
m_c \simeq \frac{2 R\sigma_c^3}{\sqrt{3}G \sigma_{\rm halo}} \sim
 \frac{2c^3R}{\sqrt{3}G\sigma_{\rm halo}} \left(\frac{l_s}{4\pi D_{\rm 
ls}}\right)^{3/2}
\sim 5\times 10^3 \msun \left( \frac{hc}{H_0D_{\rm ls}}
 \frac{l_s}{\mbox{pc}} \right)^{3/2} \frac{200\kms}{\sigma_{\rm halo}}
\frac{R}{\kpc},
\label{eq:m_c}
\eeq
where the tidal radius~(\ref{sis_m_r}) was used.  As we shall see this
estimate is fairly accurate.

\begin{figure}[t]
\centering\epsfig{figure=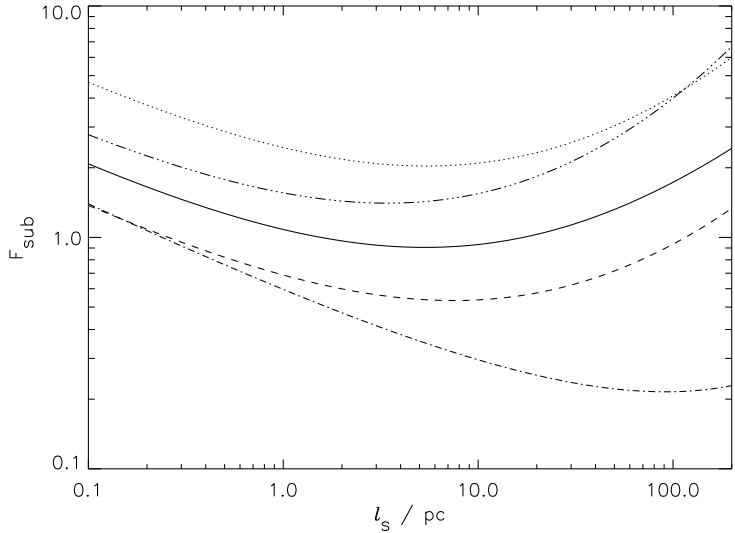,height=3in}
\vspace{0.5cm}
\caption[]{\footnotesize Some examples of the substructure lensing parameter
$F_{\rm sub}$ as a function of source size.  The solid line is for
$(f_\Sigma,\mu,R,m_{\rm max})=(0.05,9.3,1\kpc,10^{10}\msun)$ where
$f_\Sigma$ in this case is the fraction of mass in subclumps with $m>m_c$.
Parameterized in the same way - {\it dotted}
$(0.1,7.3,1\kpc,10^9\msun)$, {\it dashed} $(0.03,5.0,1\kpc,10^9\msun)$,
{\it dot--dashed} $(0.03,3.0,5\kpc,10^9\msun)$, {\it dot--dot--dashed}
$(0.05,9.3,1\kpc,10^8\msun)$.  In all cases $z_s=3$, $z_l=1$, the
subclump mass function is $\propto m^{-2}$, and we assumed a flat cosmology 
with $\Omega_M=1$.
}
\label{fig:lens_parameter}
\end{figure}

If subclumps with $m>m_c$ are rare because the mass fraction in this
mass range is small, then it will be unlikely for images to be affected.  We
can introduce a figure of merit that gauges the effectiveness of
substructure to influence images by considering the average number of
satellites that intersect an image.  An idealized circular
image will have a radius of $l_sD_l\sqrt{|\mu|}/D_s$ on the lens
plane.  If we add this radius to the radius of the subclump we can find
the area in which a subclump will influence the image.  The average number
of subclumps in this region is,
\begin{eqnarray}
F_{\rm sub}= \pi \int^{m_{\rm max}}_{m_c} dm'~
\left[ r(m',\kappa) + \frac{l_sD_l}{D_s}\sqrt{|\mu|} \right]^2 \frac{d\eta}{dm}(m').
\end{eqnarray}
Roughly, if $F_{\rm sub} \simgt 1$ substructure will affect almost every
image.  How well $F_{\rm sub}$ characterizes the lensing properties
needs to be investigated with numerical simulations, but we can glean
some expectations from Figure~\ref{fig:lens_parameter} where $F_{\rm sub}$
is plotted for some example sets of satellite parameters. Clearly, only a small
fraction of the mass needs be in subclumps, $\sim 1\%$, for them to have a
significant influence on images. Generally,  when the source is small
substructure is important because $m_c$ is small and there are many
available lenses. As $l_s$ increases
$F_{\rm sub}$ decreases until the source gets large enough that it becomes
probable for one of the larger satellites to lie close to the image.
$F_{\rm sub}$ would not increase again for large $l_s$ as it does in
Figure~\ref{fig:lens_parameter} if the mass function were more
bottom--heavy than $m^{-2}$.  When the grow subclumps larger while
keeping their masses fixed the minimum of $F_{\rm sub}$ is smaller and
moves to larger $l_s$; the parameter $R$ is used here to normalize the
size distribution through equation~(\ref{sis_m_r}).  $F_{\rm sub}$ does
not increase indefinitely with $l_s$ because at some point $m_{\rm
max}\simeq m_c(l_s)$, and satellites cease to have much of an effect.

The source has been taken to be circular so far, but if the source is
highly elongated like a radio jet the likelihood of there being a
subclump nearby will be larger.  At the same
time its small width makes it susceptible to distortion by comparatively
small subclumps.  The superior resolution of radio interferometers could
make these good sources for observing substructure lensing if the
surface brightness of the jets is high enough.

Microlensing of multiple image QSOs has already been identified through
its time dependence.  Stellar mass compact objects have lensing
timescales on the order of months.  In contrast lensing by substructure
will be nearly time independent.  The characteristic timescale in this case is
$t=\lambda_o(\sigma)/v_\perp$.  This results in
\begin{eqnarray}
t \simeq 4.1\times 10^5\, h^{-1} \left(\frac{\sigma}{\kms}\right)^2
\left(\frac{\kms}{v_\perp}\right) \left(\frac{H_0D_lD_{\rm ls}}{c\, 
D_s}\right)\yr.
\end{eqnarray}
For subclumps with $\sigma_{\rm sub}\sim 10\kms$ and $v_\perp \sim 200\kms$
this timescale is very large.  It is therefore justified to treat the
subclump as stationary.

\subsection{Numerical results}
\label{sec:Numerical-results}
In the following, different aspects of our simulation
results will be discussed which correspond to different methods for
probing substructure.  The simulations are carried out for a 
compact ($< 1 \kpc$) source at $z=3$ in an Einstein--de Sitter cosmology.
The cosmology has little effect on the results.  The subclump
mass function is $\propto m^{-2}$ in accordance with expectation from
CDM models.  A $256^2$ grid is usually used giving an estimated
numerical error in the magnification of $\delta\mu\sim A_{\rm grid}(N^2
A_{\rm source})^{-1}=10^{-5}~A_{\rm grid}/A_{\rm source}$ where the $A$'s are the
area of the grid and source.  The grid area changes according to the
subclump sizes and the background magnification, but generally an
accuracy of $\delta\mu \sim 0.01 - 0.05$ is attained.  Simulations with
higher resolution have been run to verify this accuracy estimate.

\subsubsection{Magnifications}
\label{sec:Magnifications}

\begin{figure}[t]
\vspace{-0.75in}
\centering\epsfig{figure=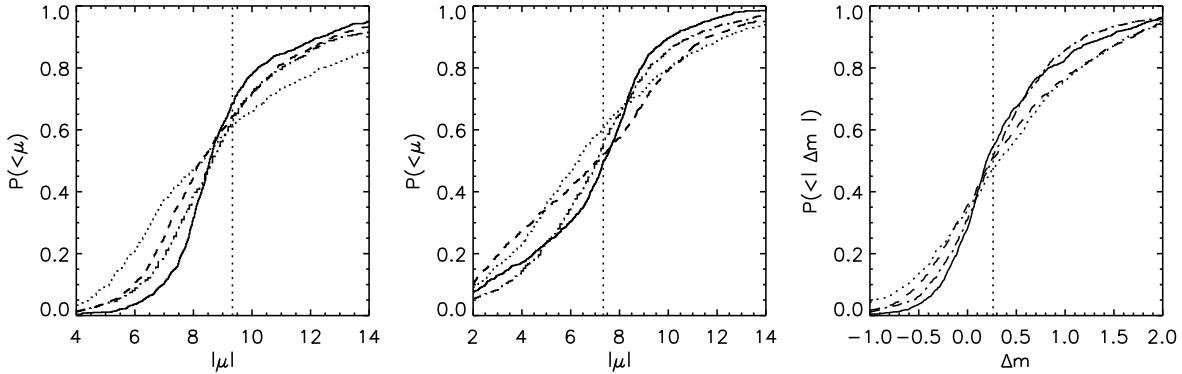,height=4.5in}
\vspace{-1.5in}
\caption[wew]{\footnotesize The cumulative magnification distributions for the two
images of a SIS lens with source position 
$y=0.12$ measured in lensing scale lengths.  The distribution of image~1
-- the primary image or outer image -- is plotted on the left.  In the center is
the distribution for image~2.  On the right the cumulative
distribution of the difference in image brightnesses measured in
magnitudes is shown. Negative values of $\Delta m$ correspond to cases where 
image~2 is brighter than image~1.  The vertical dotted
lines are the values expected for a smooth lens and an infinitely small
source.  The different curves are for different forms of substructure
and source sizes.  The {\it solid curve} assumes that $5 \%$ of the mass is in
subclumps with $10^4\msun <m< 10^8\msun$, and a 10~pc source.  {\it
Dashed line:} same except with $10\%$ of the mass in substructure.  {\it
Dash--dotted curve:} assumes $5\%$ of the mass in subclumps with
$10^3\msun <m< 10^7\msun$ and a 1~pc source. {\it Dotted curve:} same as
the dash--dotted line only with $10\%$ of the mass in substructure.}
\label{fig:mag_y12}
\end{figure}

\begin{figure}[t]
\vspace{-0.75in}
\centering\epsfig{figure=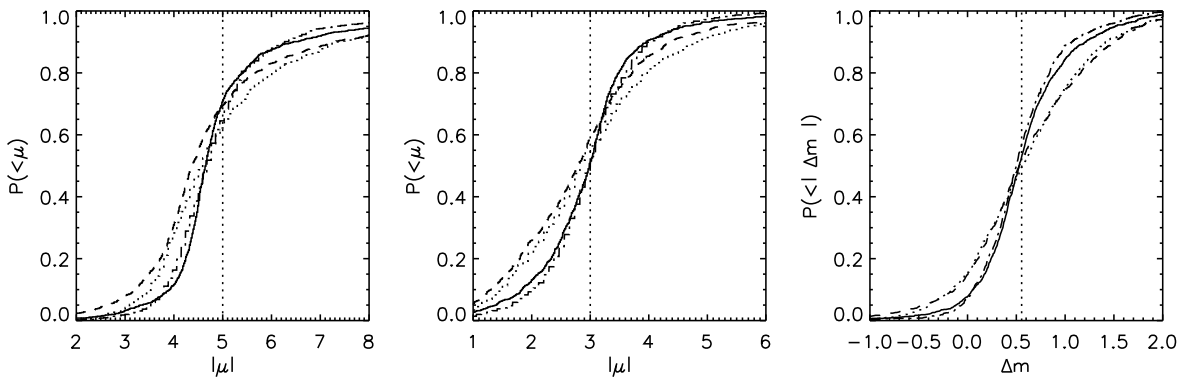,height=4.5in}
\vspace{-1.5in}
\caption[]{\footnotesize This is the same as Figure~\ref{fig:mag_y12}
only the source position is $y=0.25$.}
\label{fig:mag_y25}
\end{figure}

\begin{figure}[t]
\vspace{-1.0in}
\centering\epsfig{figure=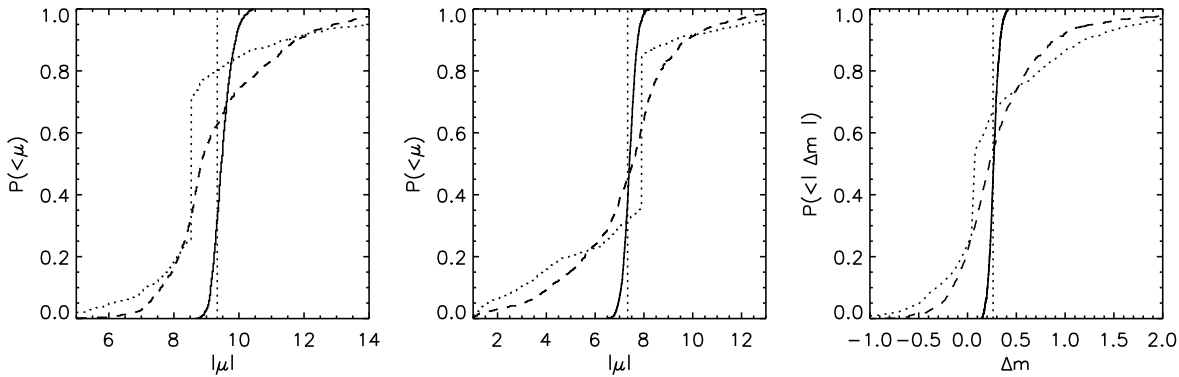,height=4.5in}
\vspace{-1.5in}
\caption[qiw]{\footnotesize The cumulative magnification distributions
for the two images of a 10 pc source with source position $y=0.12$.  The three
different curves correspond to different subclump mass ranges. {\it Solid
line:} $10^4\msun <m< 10^5\msun$. {\it Dashed line:} $10^6\msun <m< 10^7\msun$.
{\it Dotted line:} $10^7\msun <m< 10^8\msun$.  The mass fraction in subclumps,
$f_\Sigma$, is held fixed at $2\%$.  The subclumps are progressively
more influential on the lensing as they get more massive despite their
smaller number density.  The vertical sections of the dotted curves are
located at the background magnifications -- the value of the magnification
when there are no subclumps influencing the image.}
\label{fig_10pc_mass}
\end{figure}

\begin{figure}[t]
\vspace{-1.0in}
\centering\epsfig{figure=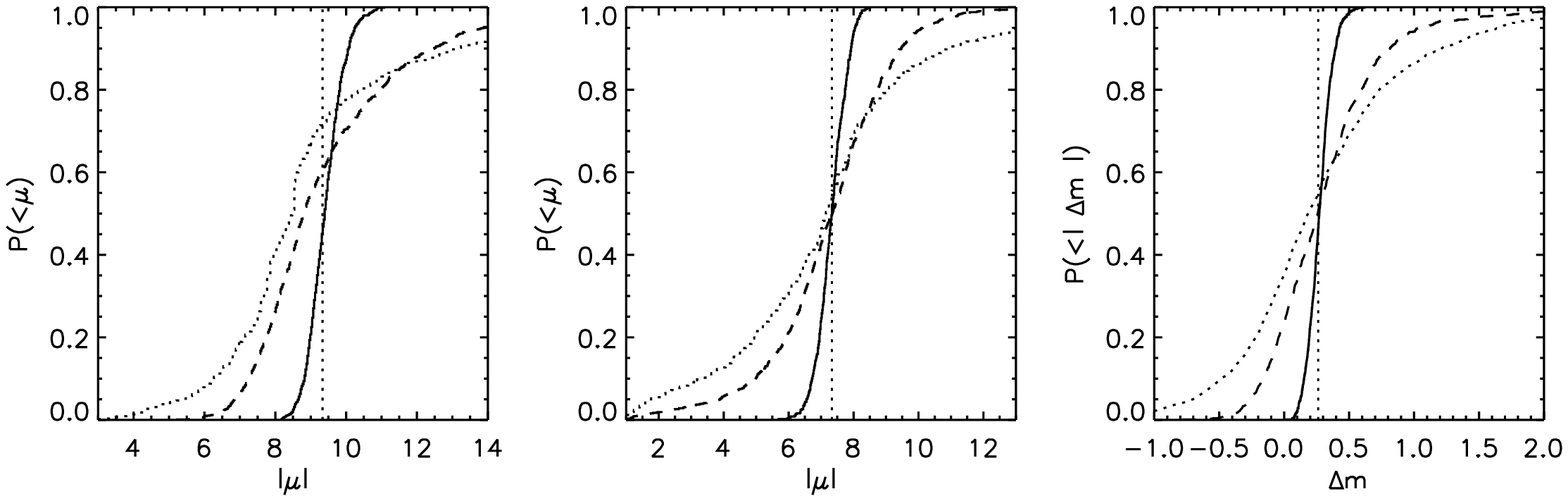,height=4.5in}
\vspace{-1.5in}
\caption[qiw]{\footnotesize Same as in Figure~\ref{fig_10pc_mass}, but
the source is 1 pc in size and the mass ranges are: {\it Solid
line:} $10^2\msun <m< 10^3\msun$. {\it Dashed line:} $10^4\msun <m< 10^5\msun$.
{\it Dotted line:} $10^5\msun <m< 10^6\msun$.}
\label{fig_1pc_mass}
\end{figure}

As will be seen in section~\ref{sec:Image-distortions} the image positions are not
significantly changed by small scale substructure.  This allows for the
possibility of constructing a smooth lens model from the image positions
and perhaps the images of the QSO's host galaxy.  The magnification
ratios predicted by this model, or family of models, can then be
compared with the observations.  This will generally require four image
systems because two image systems do not contain enough information to
constrain the smooth lens model.  For simplicity we will investigate a simple
SIS lens at $z=1$ with a velocity dispersion of $\sigma_{\rm
halo}=240\kms$; this is a high velocity dispersion for galaxies in
general, but the population of gravitational lenses tend to be highly
biased toward the largest of elliptical galaxies.  In any case, if the source
position is fixed in units of $\lambda_o(\sigma_{\rm halo})$ then
$\sigma_{\rm halo}$ only affects the subclump size distribution  -- a
smaller host halo giving smaller subclumps at the image positions.

The magnification distribution for each image is calculated by repeating
the calculation with different realizations of the substructure.
Figures~\ref{fig:mag_y12} and~\ref{fig:mag_y25} show the cumulative
distributions for two choices of the projected source position relative to
the center of the host halo.  The magnification ratios of the images --
the only directly observable quantity -- are shown on the right of each
figure.  This demonstrates that with only a few percent of the surface density
contained in substructure the deviations from the expected magnification
ratio are significant in most cases.
For comparison typical errors quoted for HST observations are $\simlt
0.05\mag$.  For the models in Figures~\ref{fig:mag_y12} only $14\% -
32\%$ of the cases, depending on the model, have ratios within 
$0.2\mag$ of the expected ratio.  For Figures~\ref{fig:mag_y25} where
the source position is further from the lens center this is $26\% -
44\%$ of the cases.  For the models in Figure~\ref{fig:mag_y12} a
significant fraction of the cases have the image that is expected to be
the brightest instead the dimmer of the two -- a sure sign that the
lens is more complicated than the lens model reflects.

To gauge the impact of substructure on lensing data consider $n$ lens
systems.  The likelihood that at least one of the
magnification ratios is discrepant by a certain amount, $m$, is $1-p_1p_2\dots
p_n$ where $p_i$ is the probability that the $i$th image pair has a
magnification ratio within $m$ of the expected ratio.  With only a
few systems a strong limit can be put on the amount of substructure as
long as the expected magnification ratios can be predicted to a certain
degree.  For example, if we were to observe 5 independent QSO image
pairs like the ones depicted in Figures~\ref{fig:mag_y12} or
\ref{fig:mag_y25}, and all of them had ratios within $0.2\mag$ of the
expected values then all the models displayed could be ruled out at 
better than $98\%$ confidence.

We expect that if $l_c^3/m^2$ is held fixed the magnification
distribution will be relatively independent of the source size and the
subclump mass range; this is because
$\lambda_o(\sigma)\propto \sigma^2 \propto m^{2/3}$ (see
\S~\ref{sec:SIS-model}).  A comparison of figures~\ref{fig:mag_y12} and
\ref{fig:mag_y25} 
is generally in agreement with this scaling relation.
Without precise knowledge of the source size the strongest
constraint would be put on $f_\Sigma$.  If the substructure masses are
reduced while keeping the source size fixed the distribution tends not
to extend to as high magnifications because there is an effective limit
to how much a small structure can magnify the large sources.
For a fixed substructure distribution, as expected from our discussion
of $F_{\rm sub}$ (section~\ref{sec:SIS-model}), the images of smaller
sources tend to be more strongly affected by substructure.  Smaller
sources are sensitive to a wider range of subclump masses.  Figures~\ref{fig_10pc_mass}
and~\ref{fig_1pc_mass} show how the lensing depends on subclump
mass.  Subclumps of mass $\simlt 10^4\msun$ do not have a significant 
influence on 10~pc sources and $m\simlt 10^2\msun$ subclumps do not affect 1~pc
sources at the position and redshift considered.  This is
in general agreement with the estimate of $m_c$ made before,
equation~(\ref{eq:m_c}).  This is also a useful test
of the simulations because as the number density of subclumps goes up
the magnification should converge to the smooth lens solution.

For the high substructure mass ranges in Figure~\ref{fig_1pc_mass} and
especially Figure~\ref{fig_10pc_mass} there is a bias toward smaller than
expected magnification ratios.  This is a result of substructure having a
different effect on different kinds of images.  To understand this
consider a small change in the convergence, $\kappa$, (surface density)
from the value predicted by the smooth model $\kappa_o$.  To first order
the change in the magnification is
\begin{eqnarray}\label{eq:mag_expansion}
\frac{\Delta\mu}{\mu_o} \simeq 2 (1-\kappa_o)\mu_o \delta\kappa ~.
\end{eqnarray}
Clumping increases the probability that an image will land on an
underdence region where $\delta\kappa<0$.  The primary image -- where
$\mu_o>1$ and $\kappa_o<1$ -- will have a smaller magnification in this
case.  For the secondary image, which is reflected with respect to the
primary, the situation is more complicated.  If $y<0.5$, like it is in our
examples, then $\mu_o<0$ and $\kappa_o<1$, which means the absolute
value of the magnification will go up when $\kappa$ goes down.  The
opposite is true when $y>0.5$ since in this case $\mu_o<0$ and
$\kappa_o>1$ for image two.   This biasing when $y<0.5$ does not seem to
have a large effect when the mass distribution is steep and extends down to
$m_c$ as in Figures~\ref{fig:mag_y12} and~\ref{fig:mag_y25}.  However,
if the substructure consists of only high mass objects -- relative to the
source size -- the bias will be significant.  This is most noticeable for
the $10^7\msun <m< 10^8\msun$ model in Figure~\ref{fig_10pc_mass}
(dotted curves) where in a large fraction of the cases the images have
the background magnification, $\mu^b$ (the vertical sections of the curves).

When confronting observations the smooth lens model
and the substructure model need to be constrained simultaneously in four
image lens systems.  This task is complicated by degeneracies in smooth
lens models which produce equally good fits to the image
positions, but different magnification ratios.  Constraining the level
of substructure in these systems requires significant further
development beyond what is presented here and will be the subject of a
future paper by one of the authors (Metcalf \& Zhao 2001).

\subsubsection{Image distortions}
\label{sec:Image-distortions}
Resolution in the visible is not high enough to directly observe
distortions to images caused by substructure in the mass range that we
are considering here.  However radio observations with sub milli--arcsecond
resolution could be used to find distortions in one image that 
are not mirrored in all the other images if the source is of the
appropriate size.  To investigate this possibility and to gain more
understanding of the lensing behavior in general we consider the
images produced in the same simulations used in the previous subsection.

\begin{figure}[t]
\epsfig{figure=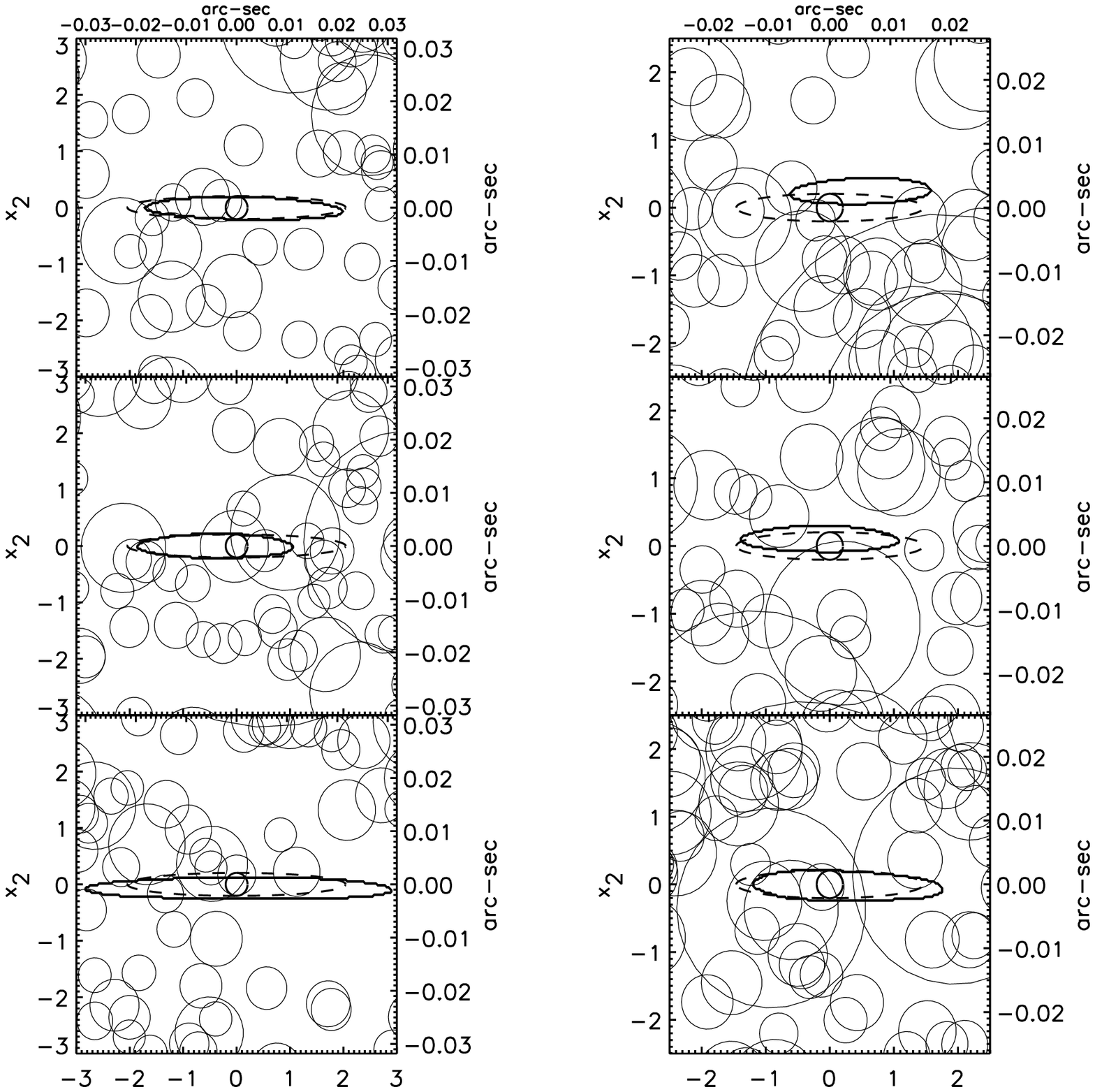,height=5.25in}
\vspace{1cm}
\caption[qiw]{\footnotesize Images of a 10~pc source at $z=3$ seen
through a lens at $z=1$.  On the left are realizations of the outer,
primary image and on the right simulations of the inner image.  The
source position is $y=0.12$.  The subclumps constitute $5\%$ of the
density and $10^4\msun <m< 10^8\msun$ with a $m^{-2}$ mass function.
For clarity only a small part of the field used in the simulations is shown
and the subclumps with $r<0.3 \lambda_o(m_{max})$ are not drawn.  This
corresponds to masses below $2.0\times 10^4\msun$.  The radii of the
circles are the tidal radii of the subclumps.  The thick
dashed curves are the shape the image would be if the lens were smooth
and the thick circles are what the source would look like if there were
no lensing.}
\label{fig:gridimage}
\end{figure}

Figure~\ref{fig:gridimage} shows representations of the images with
random realizations of substructure.  In most cases the change in
the morphology of the image is small.  The distortions are typically on
milli--arcsecond scales.  It can also be seen that the centroid of the
image is not significantly shifted.  It would take substructures with
masses of $\simgt 10^8\msun$ that are very well aligned with the image to
change the image position by a few tens of milli--arcsecond.  Such a
chance alignment will be rare even in the CDM model.

These figures also show that with the sources and substructures we are
considering the internal structure of the subclumps should not make a
great deal of difference to the lensing.  The images feel a deflection
potential that is smoothed on the scale of their own size.  Since the
images are about the same size as the subclumps in these examples how cuspy
the subclump cores are will not make a great deal of difference.

\begin{figure}[t]
\centering\epsfig{figure=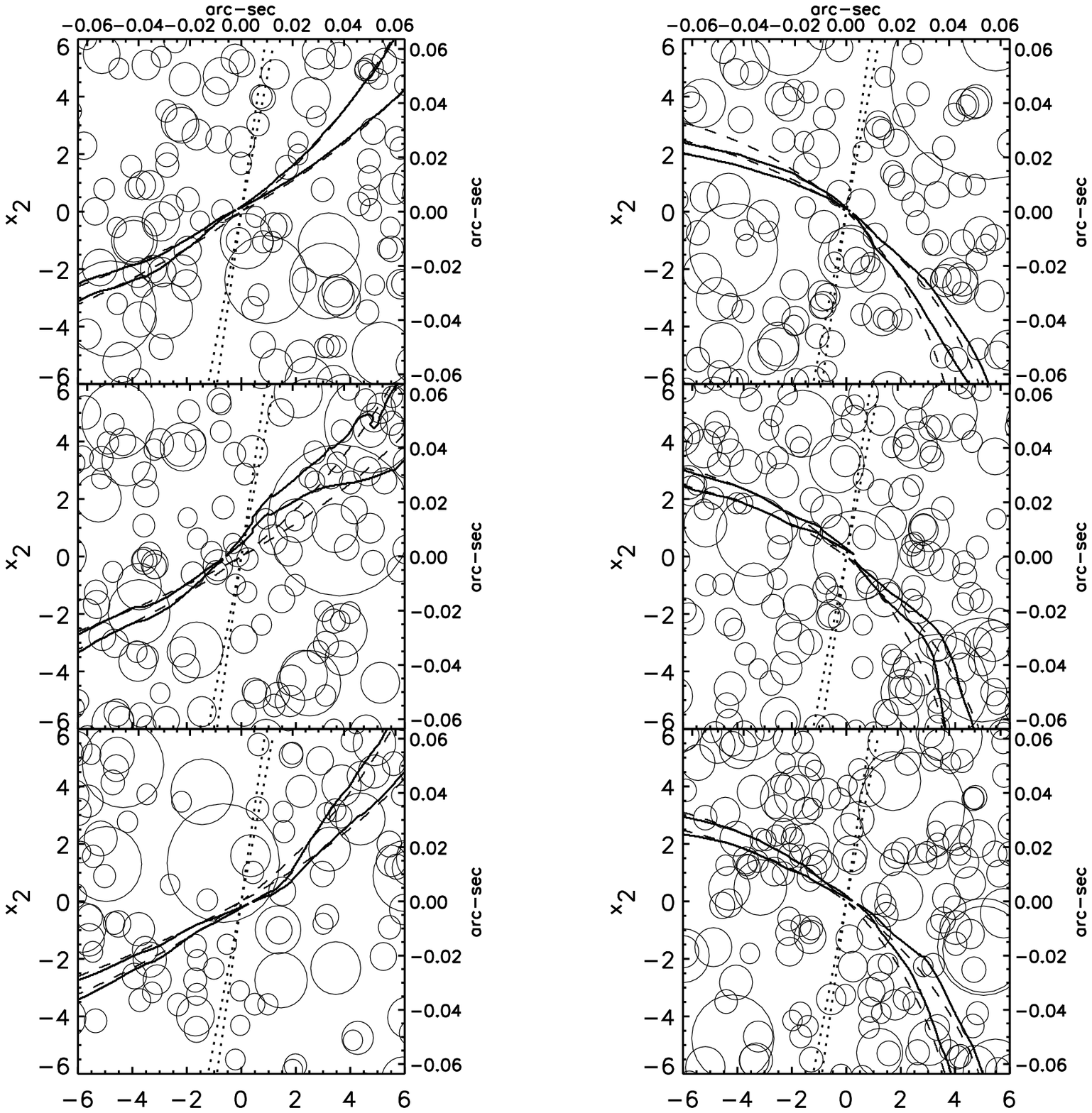,height=5.25in}
\vspace{1cm}
\caption[qiw]{\footnotesize Images of an idealized radio jet at $z=3$ seen through a
lens at $z=1$. The source of the jet is at $y=0.12$.    The
subclumps constitute $5\%$ of the density and $10^4\msun <m< 10^8\msun$
with a $m^{-2}$ mass function.  Only subclumps with
$r>0.35\lambda_o(m_{\rm max})$ ($m>3.2\times 10^4\msun$) are shown.  The unlensed jet is shown in
dotted lines.  The opening angle is $6^o$ and it is oriented at $10^o$
from the lens' radial direction that is the vertical axis here.  The
dashed lines show the image were the lens smooth.}
\label{fig:pmimage_jet}
\end{figure}

Another interesting possibility is a lensed radio jet.  Jets typically have
opening angles of a few degrees and extend over 
Mpc (see Kembhavi \& Narlika 1999 and Peterson 1997).  This source would then cover more
area on the lens plane making it more likely to come close to a large
subclump; its thinness would make distortions more pronounced.
Figure~\ref{fig:pmimage_jet} shows 
some idealized representations of what a lensed radio jet would look
like.  Since the jets are not perfectly conical in reality the
distortion caused by lensing would need to be deduced by comparing the
two images.  Some of the distortions are on $\sim 0.01$~arcsec scales.
There is at least one multiply--imaged radio jet known, Q0957 + 561.
The two images have been mapped on milli--arcsecond scales
\markcite{1988ApJ...334...42G}(Gorenstein et al. 1988) and there is
not any obvious evidence for substructure in this case although a rather
large subclump would be require to produce a noticeable effect ($\sim
10^7\msun$)\markcite{1999ApJ...520..479B}(Barkana et al. 1999).

\subsubsection{Distribution of magnification ratios}
\label{sec:Distr-magn-rati}
Another approach to take in the search for substructure is to look at
all multiply--imaged QSOs and consider the overall distribution of
magnification ratios without modeling individual cases, only the
population of halos in general.  This method has several potential
advantages besides requiring less work.  Two image QSOs 
can be used in this case, which increases the size of the sample and
generally the physical distance the images are from the center of the
lens.  The latter property means both that substructure will be more
likely to survive  and that the final constraint will be less dependent
on the smooth lens model used.  To investigate this situation we fix the
source at $z=3$, but choose a random lens redshift and impact parameter
according to the assumption of constant comoving density.  The physical
size distribution of subclumps is kept fixed and is the same as in
the previous two subsections.  In this case the lensing distribution is
independent of the distribution of lens velocity dispersions.

\begin{figure}[t]
\centering\epsfig{figure=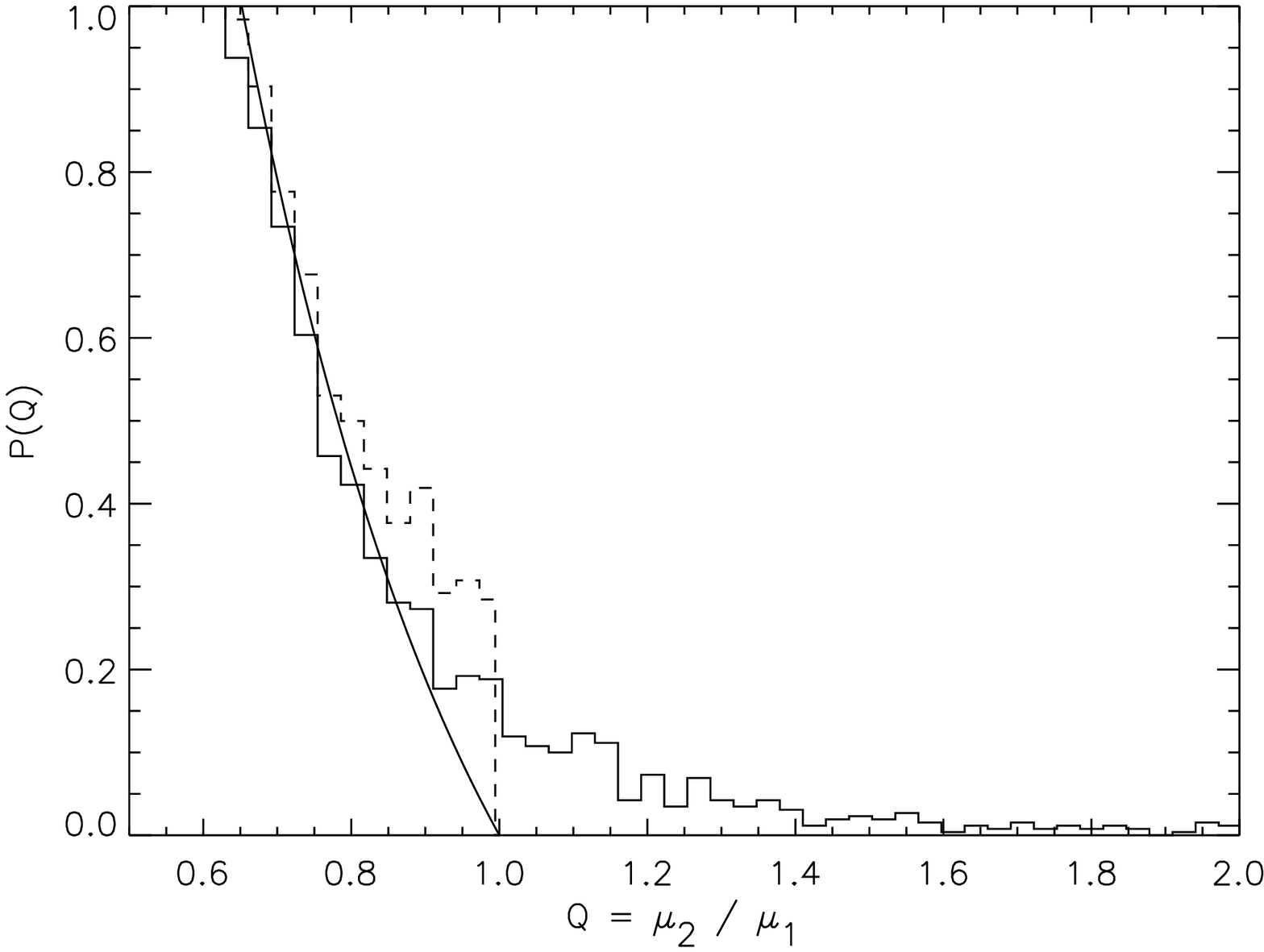,height=3in}
\caption[qiw]{\footnotesize The distribution of magnification ratios for
a 10~pc source at $z=3$.  The smooth curve is for lenses without
substructure.  The solid histogram includes the effect of subclumps. The
dashed histogram is the same except that $Q$ is the lesser of
$\mu_1/\mu_2$ and $\mu_2/\mu_1$.  The subclumps are a fixed $5\%$ of the
lens surface density, and have $10^4\msun <m< 10^8\msun$.  All three
distributions have the same integral for $Q > 2/7$.  This cutoff is
necessary because the Monte Carlo calculation cannot be efficiently
extended down to $Q=0$.  However, small $Q$ pairs are strongly selected
against in observations and the distribution is not strongly affected by
substructure in this region.  There are 10,000 realizations in the simulated
distribution.}
\label{fig:qdist}
\end{figure}

The cumulative distribution of magnification
ratios is shown in Figure~\ref{fig:qdist} along with the theoretical
prediction for smooth lenses.  The most significant effect is an over
abundance of cases with magnification ratios close to one and cases
where the image that is expected to be brighter is in fact the dimmer
image.  This is mostly the result of simple
spillover caused by extra scatter added to a steep distribution -- the
biasing discussed in section~\ref{sec:Magnifications} does not play a
big part.
Identifying the image that is expected to be brighter requires some
knowledge of the individual lenses.  In the simple case of an SIS lens
it would just require determining which image is closer to the lens; but
without this information -- which is often missing in radio surveys --
we can still consider the ratio of the dimmer image to the brighter one.
This distribution is the dashed one in Figure~\ref{fig:qdist} where it
can be seen that there is an overabundance of cases with ratios close to
one.  A proper comparison with data would require accurate modeling of
the selection effects inherent in any lens survey as well as using a
more realistic smooth lens model.  At this time the publicly available
data is very limited and does not put an interesting constraint on
substructure.  This may change in the near future with the completion of
several large, systematic QSO surveys.

\section{Discussion}
\label{sec:Discussion}

We have shown in this Paper that, if dark matter substructure exists in
the halos of galaxies and can survive to constitute a percent or more of the
surface density at small impact parameters (several kpc) -- a
possibility that is in general agreement with CDM hierarchical cosmologies, it 
will cause significant changes to the magnification ratios of multiply--imaged QSOs.  Furthermore, if the magnification ratios of just a few QSO image pairs
can be predicted from the image positions and other information to
an accuracy of a few tenths of a magnitude the existence of such
substructure can be confirmed or invalidated.  This would allow us to
probe primordial structure formation on the scales of $10^3 - 10^8\msun$.  
To confront observations more detailed modeling of individual QSO lens
systems will be needed.  This is the topic of another paper
(Metcalf \& Zhao 2001, in preparation).

Besides lens modeling there are several other complicating factors involved
in measuring the magnification ratios.  A correction for differential
extinction must be included in the visible bands.
\markcite{1999ApJ...523..617F}Falco et al. (1999) find a median
differential extinction between lensed images of $\Delta E(U-V) =
0.04-0.06 \mag$ so this is not likely to be a large impediment.  Another
issue is intrinsic variation in the source.  If the source is variable and the
time delay is not well known it is difficult to interpret the flux ratios.  
Radio observations avoid the complication of differential extinction,
but radio galaxies are often variable.  Also the lens galaxy is usually
not detected in the radio, making the constraints on the smooth lens
model weaker.  A combination of observations at different wavelengths
will be required.

Microlensing by stars in the lens galaxy is also a possible
contaminating factor if the source is too small.  The continuum emission
region of a QSO is believed to be less then $\sim 100$~au,
and is observed to be microlensed in some cases.  To avoid this a larger
source must be used.  The broad line emission region is believed to be
about a pc in size, which would make it sensitive the subclump masses
$\simgt 10^3\msun$.  There are also features in the radio emission that
are both smaller and larger than this.  They would generally need to be
resolved in order to confirm their scale.  QSO narrow line regions
probably have sizes of $\simgt 100\pc$.  The molecular line emission
region of AGN is believed to be $1-100\kpc$ in size, but it is not
certain if the CO emission from QSOs, for example, comes from a localized
region or from the host galaxy as a whole (see Kembhavi \& Narlika 1999
for a review of QSO properties).  Conclusions drawn from
lensing on the existence and mass distribution of 
substructure will thus depend on the kind of observations used and our
knowledge of the sources' internal structure.

Finally, the images could also be lensed by more familiar galactic 
substructures such as spiral arms (Mao \& Schneider 1998). 
These are expected to interfere with the constraints
on dark matter structure at a small level.  The Milky Way disk has
a surface density of $\sim 50\msun/\pc^2$ and spiral arms are believed
to have an amplitude of about $20\%$ in mass.  For a lens at $z=1$ and a
source at $z=3$ in an Einstein--de Sitter universe this corresponds to
$\kappa = 0.003 h^{-1}$.  We can use equation~(\ref{eq:mag_expansion}) to
calculate the change this would cause in the magnification, which is
$\sim 0.05\mag$ in the case of the most magnified image discussed in
section~\ref{sec:Examples}.  The dark matter subclumps considered here
have more influence on the magnification because of their comparatively
high mass density.  Globular clusters are unlikely to cause significant
changes in the magnifications because they do not contain a significant
fraction of the halo mass ($\sim 10^{-4}$ of the Milky Way halo) as
pointed out by Mao \& Schneider (1998).

\acknowledgments
\noindent R.B.M. would like to thank H. Zhao for many constructive
discussions.  We would like to thank the anonymous referee for
constructive criticism.  Support for this work was provided by NASA
through ATP grant NAG5--4236 and grant AR--06337.10-94A from the Space
Telescope Science Institute (P.M.).


\begin{thebibliography}{}
\bibitem[{Barkana}, {Leh{\'a}r}, {Falco}, {Grogin}, {Keeton}, \&
{Shapiro} 1999]{1999ApJ...520..479B} {Barkana}, R., {Leh{\'a}r}, J.,
{Falco}, E.\ E., {Grogin}, N.\ A., {Keeton}, C.\ R., {Shapiro}, I.\ I. 1999, \apj, 520, 479

\bibitem[Binney \& Tremaine 1987]{BinneyAndTremaine}
Binney, J. \& Tremaine, S. 1987, Galactic Dynamics (Princeton, New Jersey:  Princeton University Press)

\bibitem[{Blitz}, {Spergel}, {Teuben}, {Hartmann}, \&  {Burton} 1999]{1999ApJ...514..818B}
{Blitz}, L., {Spergel}, D.~N., {Teuben}, P.~J., {Hartmann}, D., \& {Burton},  W.~B. 1999, \apj, 514, 818

\bibitem{BOT} Bode, P., Ostriker, J. P., \& Turok, N. 2000, preprint
(astro-ph/0010389)

\bibitem[Bullock, Kolatt, Sigad, Somerville, Kravtsov,  Klypin, Primack, \& Dekel 1999]{astro-ph/9908159}
Bullock, J., Kolatt, T., Sigad, Y., Somerville, R., Kravtsov, A., Klypin, A.,  Primack, J., \& Dekel, A. 1999, astro-ph/9908159

\bibitem[{Bullock}, {Kravtsov}, \&  {Weinberg} 2000]{2000ApJ...539..517B}
{Bullock}, J.~S., {Kravtsov}, A.~V., \& {Weinberg}, D.~H. 2000, \apj, 539, 517

\bibitem[{Carr} \& {Lacey} 1987]{1987ApJ...316...23C} {Carr}, B. J. \& {Lacey}, C.\ G. 1987,
\apj, 316, 23.

\bibitem{Col00} Colin, P., Avila--Reese, V., \& Valenzuela, O. 2000, 
ApJ, 542, 622
  
\bibitem[{Dekel} \& {Silk} 1986]{1986ApJ...303...39D}
{Dekel}, A. \& {Silk}, J. 1986, \apj, 303, 39

\bibitem{1999ApJ...523..617F} Falco, E.E., Impey, C.D., Kochanek, C.S., 
Leh\' ar, J., McLeod, B.A., Rix, H., Keeton, C.R., Mu\~ noz, J.A. \&
Peng, C.Y. 1999, \apj, 523 617.

\bibitem{FP} Flores, R. A., \& Primack, J. R. 1996, ApJ, 457, L5

\bibitem[{Ghigna}, {Moore}, {Governato}, {Lake},  {Quinn}, \& {Stadel} 1998]{1998MNRAS.300..146G}
{Ghigna}, S., {Moore}, B., {Governato}, F., {Lake}, G., {Quinn}, T., \&  {Stadel}, J. 1998, \mnras, 300, 146

\bibitem[{Gorenstein}, {Cohen}, {Shapiro}, {Rogers},
{Bonometti}, {Falco}, {Bartel} \& {Marcaide} 1988]{1988ApJ...334...42G}
{Gorenstein}, M.\ V., {Cohen}, N.\ L., {Shapiro}, I.\ I., {Rogers}, A.\
E.\ E., {Bonometti}, R.\ J., {Falco}, E.\ E., {Bartel}, N. \&
{Marcaide}, J.\ M. 1988, \apj, 334, 42

\bibitem[{Johnston}, {Spergel}, \&  {Hernquist} 1995]{1995ApJ...451..598J}
{Johnston}, K.~V., {Spergel}, D.~N., \& {Hernquist}, L. 1995, \apj, 451, 598

\bibitem[Kamionkowski \& Liddle 2000]{2000PRL.Kamionkowski}
Kamionkowski, M. \& Liddle, A. 2000, PRL, 84, 4525

\bibitem{KM01} Keeton, C. R., \& Madau, P. 2001, ApJ, 549, L25

\bibitem{KM01} Kembhavi, A.K., \& Narlika, J.V., 1999, Quasars and
Active Galactic Nuclei: An Introduction (Cambridge University Press)
 
\bibitem[{Klypin}, {Gottl{\"o}ber},  {Kravtsov}, \& {Khokhlov} 1999a]{1999ApJ...516..530K}
{Klypin}, A., {Gottl{\"o}ber}, S., {Kravtsov}, A.~V., \& {Khokhlov}, A.~M.  1999a, \apj, 516, 530

\bibitem{Kly01} Klypin, A., Kravtsov, A.~V., Bullock, J. S., \& 
Primack, J. R. 2001, ApJ, in press (astro-ph/0006343)

\bibitem[{Klypin}, {Kravtsov},  {Valenzuela}, \& {Prada} 1999b]{1999ApJ...522...82K}
{Klypin}, A., {Kravtsov}, A.~V., {Valenzuela}, O., \& {Prada}, F.  1999b, \apj, 522, 82

\bibitem{LiO} Li, L.--X., \& Ostriker, J. P. 2001, preprint (astro-ph/0010432) 

\bibitem{MS98} Mao, S., \& Schneider, P. 1998, MNRAS, 295, 587

\bibitem{Mateo} Mateo, M. 1998, ARAA, 36, 435

\bibitem{Moore01} Moore, B. 2001, in 20th Texas Symposium on Relativistic
Astrophysics and Cosmology (AIP Conf. Series), in press (astro-ph/0103100) 

\bibitem[{Moore}, {Ghigna}, {Governato},  {Lake}, {Quinn}, {Stadel}, \& {Tozzi} 1999a]{1999ApJ...524L..19M}
{Moore}, B., {Ghigna}, S., {Governato}, F., {Lake}, G., {Quinn}, T., {Stadel},  J., \& {Tozzi}, P. 1999a, \apjl, 524, L19

\bibitem[{Moore}, {Katz}, \&  {Lake} 1996]{1996ApJ...457..455M}
{Moore}, B., {Katz}, N., \& {Lake}, G. 1996, \apj, 457, 455

\bibitem[{Moore}, {Quinn}, {Governato},  {Stadel}, \& {Lake} 1999b]{1999MNRAS.310.1147M}
{Moore}, B., {Quinn}, T., {Governato}, F., {Stadel}, J., \& {Lake}, G.
1999b, \mnras, 310, 1147

\bibitem[{Moore} \& {Silk} 1995]{1995ApJ...442L...5M} {Moore}, B.  and
{Silk}, J. 1995, \apjl, 442, L5.

\bibitem[{Navarro}, {Frenk}, \&  {White} 1997]{1997ApJ...490..493N}
{Navarro}, J.~F., {Frenk}, C.~S., \& {White}, S. D.~M. 1997, \apj, 490, 493

\bibitem{Peterson97} Peterson, B.M., 1997, An Introduction to Active
Galaxtic Nuclei (Cambridge University Press)

\bibitem[{Schneider}, {Ehlers}, \& {Falco} 1992]{SEF92}
{Schneider}, P., {Ehlers}, J., \& {Falco}, E.~E. 1992, Gravitational Lenses  (Springer-Verlag)

\bibitem{SK} Sellwood, J., \& Kosowsky, A. 2001, in Gas and Galaxy Evolution 
(ASP Conf. Series), in press (astro-ph/0009074) 

\bibitem[{Spitzer} 1958]{1958ApJ...127...17S}
{Spitzer}, L.~J. 1958, \apj, 127, 17

\bibitem[{Tormen}, {Diaferio}, \&  {Syer} 1998]{1998MNRAS.299..728T}
{Tormen}, G., {Diaferio}, A., \& {Syer}, D. 1998, \mnras, 299, 728

\bibitem{vdB01} van den Bosch, F. C., \& Swaters, R. A. 2001, MNRAS, 
in press (astro-ph/0006048)

\end{thebibliography}
\end{document}